\documentclass[journal]{IEEEtran}
\usepackage{subfigure}
\usepackage{cite}
\usepackage{hyperref}
\usepackage{csquotes}
\usepackage{balance}
\usepackage{color}
\usepackage{subfigure}
\usepackage{graphicx}
\usepackage{amssymb}
\usepackage{tabularx}
\usepackage{lipsum}
\usepackage{amsmath}

\hypersetup{pdfstartview={XYZ null null 1.00}}

\begin{document}

\title{Get your Foes Fooled: Proximal Gradient Split Learning for Defense against Model Inversion Attacks on IoMT data}

\author{ Sunder Ali Khowaja, \IEEEmembership{Senior Member IEEE}, Ik Hyun Lee*, Kapal Dev, \IEEEmembership{Senior Member IEEE}, Muhammad Aslam Jarwar, 
\IEEEmembership{Senior Member IEEE},
and Nawab Muhammad Faseeh Qureshi*, \IEEEmembership{Senior Member IEEE} % <-this % stops a space
%\thanks{$^\dagger$Joint first authors, with equal contributions to this paper}
\thanks{*Corresponding authors}%:
\thanks{Sunder Ali Khowaja is with Department of Mechatronics Engineering, Tech University of Korea, Republic of Korea, and Department of Telecommunication Engineering, University of Sindh, Pakistan. Email: sandar.ali@usindh.edu.pkm sunder.ali@ieee.org}% <-this % stops a space
\thanks{Ik Hyun Lee is with Department of Mechatronics Engineering, Tech University of Korea, Republic of Korea. Email: ihlee@kpu.ac.kr}% <-this % stops a space
%\thanks{Sahil Garg is with the Electrical Engineering Department, \'Ecole de technologie sup\'erieure, Montr\'eal, QC H3C 1K3, Canada (e-mail: sahil.garg@ieee.org)}
\thanks{Kapal Dev is with the Departmet of Computer Science, Munster Technological University, Ireland and Department of institute of intelligent systems, University of Johannesburg, South Africa, e-mail:kapal.dev@ieee.org}% <-this % stops a space 
%\thanks{Sahil Garg is with the Electrical Engineering Department, \'Ecole de technologie sup\'erieure, Montr\'eal, QC H3C 1K3, Canada (e-mail: sahil.garg@ieee.org)}
%\thanks{Shahid Mumtaz is with the Instituto de Telecomunicações, P-3810-193 AVEIRO – PORTUGAL (e-mail: smumtaz@av.it.pt)}
\thanks{Muhammad Aslam Jarwar is with Department of Computing, Sheffield Hallam University, Sheffield, UK (e-mail: a.jarwar@shu.ac.uk)}
\thanks{Nawab Muhammad Faseeh Qureshi is with Department of Computer Education, Sungkyunkwan University, Republic of Korea  (e-mail: faseeh@skku.edu)}
}


\markboth{IEEE Transactions on Network Science and Engineering}%
{Shell \MakeLowercase{\textit{et al.}}: Bare Demo of IEEEtran.cls for IEEE Journals}

\maketitle

\begin{abstract}
The past decade has seen a rapid adoption of Artificial Intelligence (AI), specifically the deep learning networks, in Internet of Medical Things (IoMT) ecosystem. However, it has been shown recently that the deep learning networks can be exploited by adversarial attacks that not only make IoMT vulnerable to the data theft but also to the manipulation of medical diagnosis. The existing studies consider adding noise to the raw IoMT data or model parameters which not only reduces the overall performance concerning medical inferences but also is ineffective to the likes of deep leakage from gradients method. In this work, we propose proximal gradient split learning (PSGL) method for defense against the model inversion attacks. The proposed method intentionally attacks the IoMT data when undergoing the deep neural network training process at client side. We propose the use of proximal gradient method to recover gradient maps and a decision-level fusion strategy to improve the recognition performance. Extensive analysis show that the PGSL not only provides effective defense mechanism against the model inversion attacks but also helps in improving the recognition performance on publicly available datasets. We report 14.0$\%$, 17.9$\%$, and 36.9$\%$ gains in accuracy over reconstructed and adversarial attacked images, respectively.  
\end{abstract}
\IEEEpeerreviewmaketitle
\section{Introduction}
\begin{figure*}[!htbp] %!t
\centering
\includegraphics[width=\linewidth]{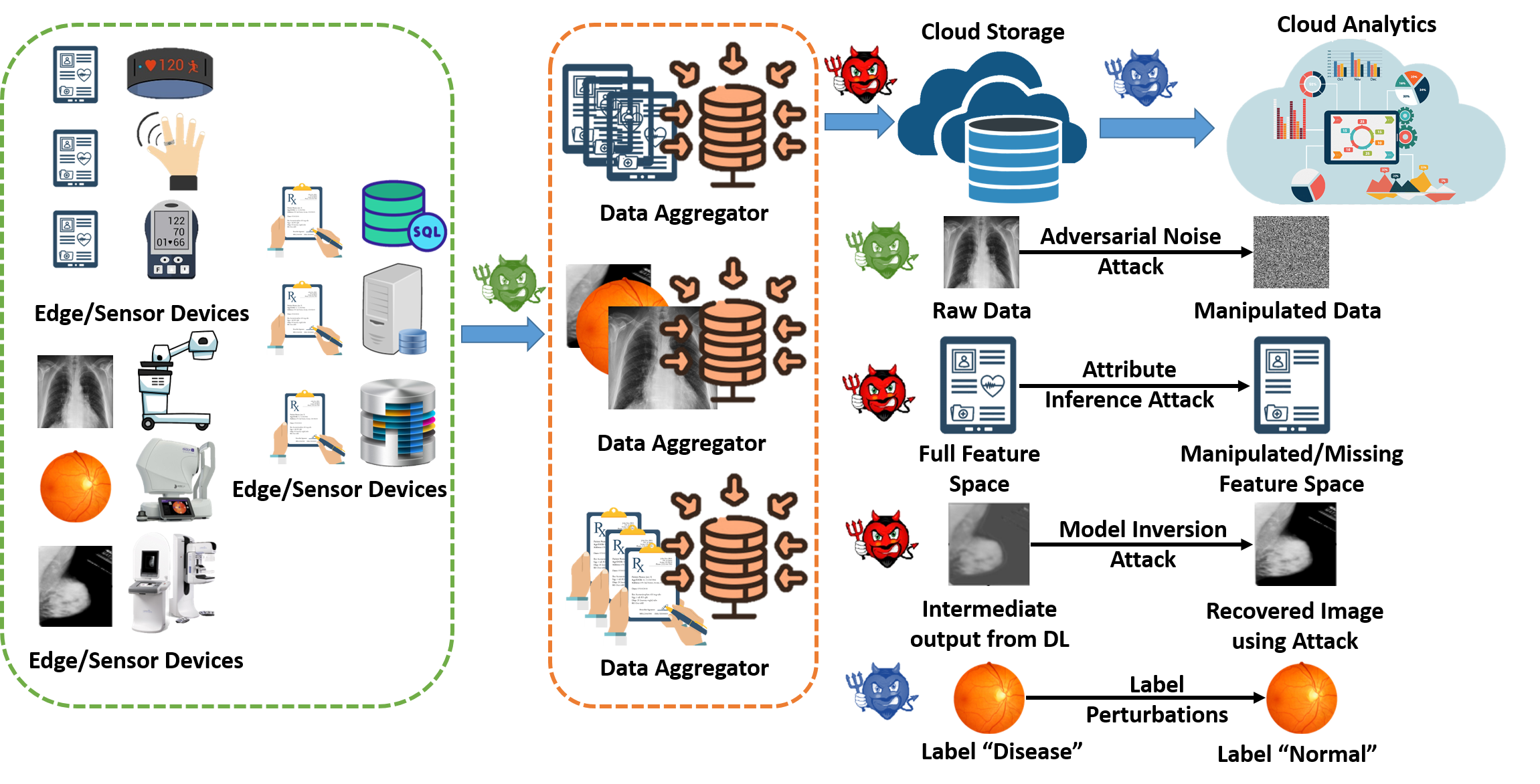}
\caption{Wide array of attacks in DL based IoMT Ecosystem}
\label{fig:1}
\end{figure*}

Modish growth in information, communication, and computing technologies have given rise to Deep learning (DL) and Internet of Things (IoT). Both computing paradigms, when combined, cater to a vast array of business requirements, technological benefits, and critical domain applications including industry, energy, transport, and healthcare sectors. The IoT covers the spectrum of data generation and collection from ubiquitous devices, while underlying intelligence and automation lies on the shoulders of DL techniques. Over the years, the use of DL in IoT ecosystem has recorded unprecedented achievements by deriving automated inferences that were too complicated for the conventional paradigms \cite{Khalil2021}.\\
The amalgamation of DL and IoT has been gaining a lot of interest in healthcare field lately, specifically by associated practitioners and researchers. Medical data comprise various modalities such as pathology test results, COVID-19 results, biomedical images, and electronic health records. The corresponding medical data acquired from IoT devices is often referred to as Internet of Medical Things (IoMT) data. Some systems that are popular and being used in the medical field are but not limited to: 1) DL based breast cancer risk prediction from mammograms \cite{Yala2019}; 2) Detection of macular edema and diabetic retinopathy using DL and retinal fundus images \cite{Li2021}; 3) Pattern detection from electronic health records using DL to determine risk factors and health trends \cite{Landi2020}.\\
Although the analytical results show drastic improvements, the issues concerning privacy of IoMT data remains at large. Medical institutions and IoMT data intrinsically hold a lot of individual's private information sch as age, gender, home address, drug usage patterns, medical history, medical test results, and so forth. The huge amount of sensitive information has attracted lots of black hats for scoring monetary, commercial, and political gains through the exploitation of IoMT data. Such information can be either leaked or intercepted when passed to the DL model for training or inferencing, respectively \cite{Wu2021}. Last couple of years have witnessed a drastic increase in attacks concerning IoMT data or DL networks. The two most common attacks that exploit personal information from IoMT data are attribute inference and model inversion attacks, respectively. The former uses partial data and a trained DL model to infer the missing piece of information, while the latter attacks intermediate layers of the trained DL models and uses the feature maps to recover the data itself \cite{Wu2021, Yuan2019}. The arousal of such attacks has hindered the hospitals' and patients' willingness to share the IoMT data and to use the DL for automated healthcare services. The lack of trust and data availability has slowed the research progress, accordingly. Therefore, it is essential to mitigate the attacks on IoMT data and develop necessary defenses for model inversion attacks. An example of adversarial attacks performed in the context of DL based IoMT data at different layers is shown in Figure \ref{fig:1}. The data can face adversarial attacks at edge device layer, aggregation layer, cloud storage or cloud analytics layer, accordingly. However, each of the attack impart different characteristics on the raw data or the derived inference. \\
Existing works have developed defense mechanisms by adding label perturbations, model perturbations, or adding noise at the data input level. The label perturbations manipulate the class probabilities that not only affects the performance of decision analysis but also compromises the protection to raw data itself. Model perturbations adds noise to the model parameters which does not provide a reasonable defense mechanism against the inversion attacks, thus, raw data can be created by employing simple pre-processing techniques. Split networks \cite{Vepa2018} were proposed to preserve the raw data privacy, but it is still susceptible to model inversion and attribute inference attacks. The work in \cite{Titcombe2021} also proposes split networks but with a random noise addition layer to combat with model inversion attacks. The aforementioned works provide a basis of using split networks to mitigate the model inversion attacks, but does not achieve desired results in terms of recognition performance. Furthermore, as revealed in our analysis, the aforementioned works do not provide a suitable defense mechanism against the model inversion attacks. In this work, we propose the proximal gradient split learning (PGSL) for prevention against model inversion attacks. We adopt the proximal gradient method and modify it as per the proposed network's requirement. The proximal gradient method is opted in this study due to its characteristics that include bounded perturbation resilience, strong convergence, and ability to handle non-smooth convex optimization problems. The network initiates an intentional one- and few- pixel attack to the input data, followed by a split deep neural network. We employ proximal gradients to reconstruct the data into its original form at the server side of the split networks. To the best of our knowledge, proximal gradients has not been explored for preserving data privacy within the training process. The contributions of this work are summarized below: 
\begin{itemize}
    \item Initiation of adversarial attack on IoMT data for improving resilience.
    \item Proximal Gradient Split Learning for training the network with adversarial samples.
    \item Late fusion strategy for improving the predictive performance on adversarial samples.
    \item Experimental analysis for validating the effectiveness of PGSL network.
\end{itemize}

The rest of the paper is structured is follows: Section \ref{sec:rel} provides a brief literature review of the existing works. Section \ref{sec:threat}  presents the threat model. Section \ref{sec:methodlogy}  provides the details regarding the proposed PGSL. Section \ref{sec:exp}  presents the experimental setup and analysis. Section \ref{sec:disc}  presents the insights, discuss the implications and limitations of the proposed work. Section \ref{sec:con}  concludes the study along with potential future works.

\section{Related Works} \label{sec:rel}
Over the years, the adoption of DL techniques for IoMT based critical and real-world services have been increased manifold. However, in recent years, research studies have exploited several vulnerabilities associated with DL in the form of adversarial perturbations. The adversarial attack was first pioneered in \cite{Szegedy2014}, that used limited-memory Broyden-Fletcher-Goldfarb-Shanno (L-BGFS) method by searching minimal distorted space to generate adversarial samples. Similarly, Goodfellow et al. \cite{Goodfellow2014} generated adversarial examples by using fast gradient sign method to perform one-step update for each pixel in the direction of gradients. Several other studies including DeepFool \cite{Dezfooli2016} and universal adversarial perturbations (UAP) \cite{Dezfooli2017} have exploited the DL techniques for its susceptibility to adversarial perturbations.\\
The domain of IoMT is mainly threatened by the data leakage and privacy attacks that target inference and training data, accordingly. The most common types of attack on IoMT data include adversarial noise attack, model encoding attack, attribute inference attack, and model inversion attack \cite{Wu2021}. 

Adversarial noise attack can be performed on a single pixel or multiple pixels, accordingly. Su et al. \cite{Su2019} proposed the adversarial noise attack by corrupting a single pixel within the specific window size to degrade the DL performance. The aforementioned study shows that 16.04\% of CIFAR10 and 67.97\% of ImageNet dataset can be attacked by manipulating a single pixel value, thus causing DL to infer the wrong label. Existing studies have tried to counter this attack through patch selection denoiser \cite{Denoiser2019}, image reconstruction \cite{Liu2020}, and adversarial detection networks \cite{Shah2020} but either they are too computationally complex or add blur artifacts to the original image. It is apparent by the studies that when corrupted by one/few pixel attacks, it is difficult to not only recover the image but also the information concerning the inference.

Attribute inference attack refers to the attacking of sensitive or prominent attributes that could either help in reconstructing the raw data or downgrade the predictive performance of the DL model. Attribute inference attack and adversarial noise attack can be similar in the case of medical images, as they both strive to corrupt the raw data itself. The study \cite{Mei2020} propose the use of attribute inference attacks to increase the risk of data theft and privacy concerning  convolutional neural network (CNN) models.

Model inversion attacks are mostly focused on reconstructing the input data from  compromised model parameters or intermediary outputs of DL methods. The studies concerning model inversion attacks can be classified into two categories. The first category refers to the set of studies that propose the use of model inversion attack to highlight the vulnerability of data using trained models, and the second category refers to the set of studies that propose defense against such attacks. The proposed method resides in the latter category. The study \cite{Fred2015} proposed the model inversion attack for recovering input images from intermediate outputs using softmax model's confidence scores. The study in \cite{Hitaj2017} proposed the use of generative adversarial networks (GANs) and collaborative training system to recover the input image from intermediary DL architecture layers. It has also been suggested by the studies \cite{Wu2021, He2019} that the reconstruction of intermediary outputs from DL architectures works better when extracted from initial layers, as they tend to have a structural similarity with the input data. Chen et al. \cite{Chen2021}, proposed the use of knowledge-enriched distributional model inversion attacks to improve the attack's success rate and add generalization across multiple datasets. Subbanna et al. \cite{Subbanna2021} performs an analytical review for the effect of model inversion attacks on medical image segmentation task using U-Net and SegNet, respectively. The study in \cite{Zhu2020} proposed the method deep leakage from gradients that proposed a differentiable model which matches the weight gradients with that of the trained model in order to reconstruct the data. The study proposed by \cite{Zhao2020} improved the weight matching algorithm for deep leakage from gradients to enhance the reconstruction performance. Similarly, the study in \cite{Wainakh2021} also proposed the reconstruction of data while matching the gradient information. In addition, the study also uses auxiliary information in order to improve the reconstruction performance.\\
In this regard, NoPeekNN \cite{Vepa2019} limited the distance correlation between the intermediate tensors and the input data during the training process of splitNN. The method was specifically designed for autoencoders to limit the reconstruction of the input data, but has not been applied or tested concerning model inversion attacks. The works \cite{Wu2021, Titcombe2021, Vepa2018} proposed the use of noise addition to the intermediate tensors, which eventually helps to cope with model inversion attacks but fails to achieve the model's accuracy.  The study does not provide any defense against the model, rather improves its ability to distill knowledge from the trained model. Titcombe et al. \cite{Titcombe2021} used noise to corrupt the intermediate data and used NoPeekNN for defense against model inversion attacks. However, the work ignores the attacks that could be initiated at the input part of the client side concerning SplitNN. Wang et al. \cite{Wang2021}, proposed the use of mutual information regularization to cope with the model inversion attacks. The method has been validated using different shallow learning methods and face dataset. Furthermore, the method uses end-to-end learning that does not deal with the information availability constraint in contrast to split networks. In this work, we propose the defense against model inversion attacks on IoMT data while dealing with split network constraints of having the network trained and inferred on server and client sides, respectively. The proposed method intentionally initiates the one/few pixel attacks in order to keep the input data safe on the client side, the intermediate output from the attacked image is then sent to the server side. We use proximal gradient method to recover the image on the client side and use late fusion technique to not only deal with model inversion attack but also with the improvement of model's accuracy.
\section{Threat Model}\label{sec:threat}
As depicted in Figure \ref{fig:1}, the attacks on the medical data can yield severe implications for not only the inference system, but for the user/patient as well. The threat model in this work considers an arbitrary number of clients that are responsible for training a part of the network and a computation server that carries on the training process on the server side. We presume that one party intends to fetch the data from other clients using model inversion attack.The process of attack is defined as follows: 1) The attackers gets their hands on data sent from database to the DL network and the intermediate feature maps from the model segment at client side; and 2) A model is trained by the attacker to reconstruct the raw data from intermediary feature maps from the client side. The aforementioned process is mainly categorized as a black-box attack \cite{Titcombe2021}. This study also assumes that there is only a single computation server and that a third party orchestrates the model training process. \\
This study only considers the intermediate data fetched from the client side or from the server's input side for model inversion attacks. This study does not take into account the data collection during the training and susceptibility of split neural networks towards Sybil attacks or membership inference attacks. Furthermore, this study also not covers the spectrum of white-box model inversion attacks, accordingly. Due to the wide spectrum of threat models, we have limited our study to black-box attack, as it helps in performing a fair comparison with existing studies.

\section{Proposed Methodology} \label{sec:methodlogy}
\begin{figure*}[!htbp] %!t
\centering
\includegraphics[width=0.8\linewidth]{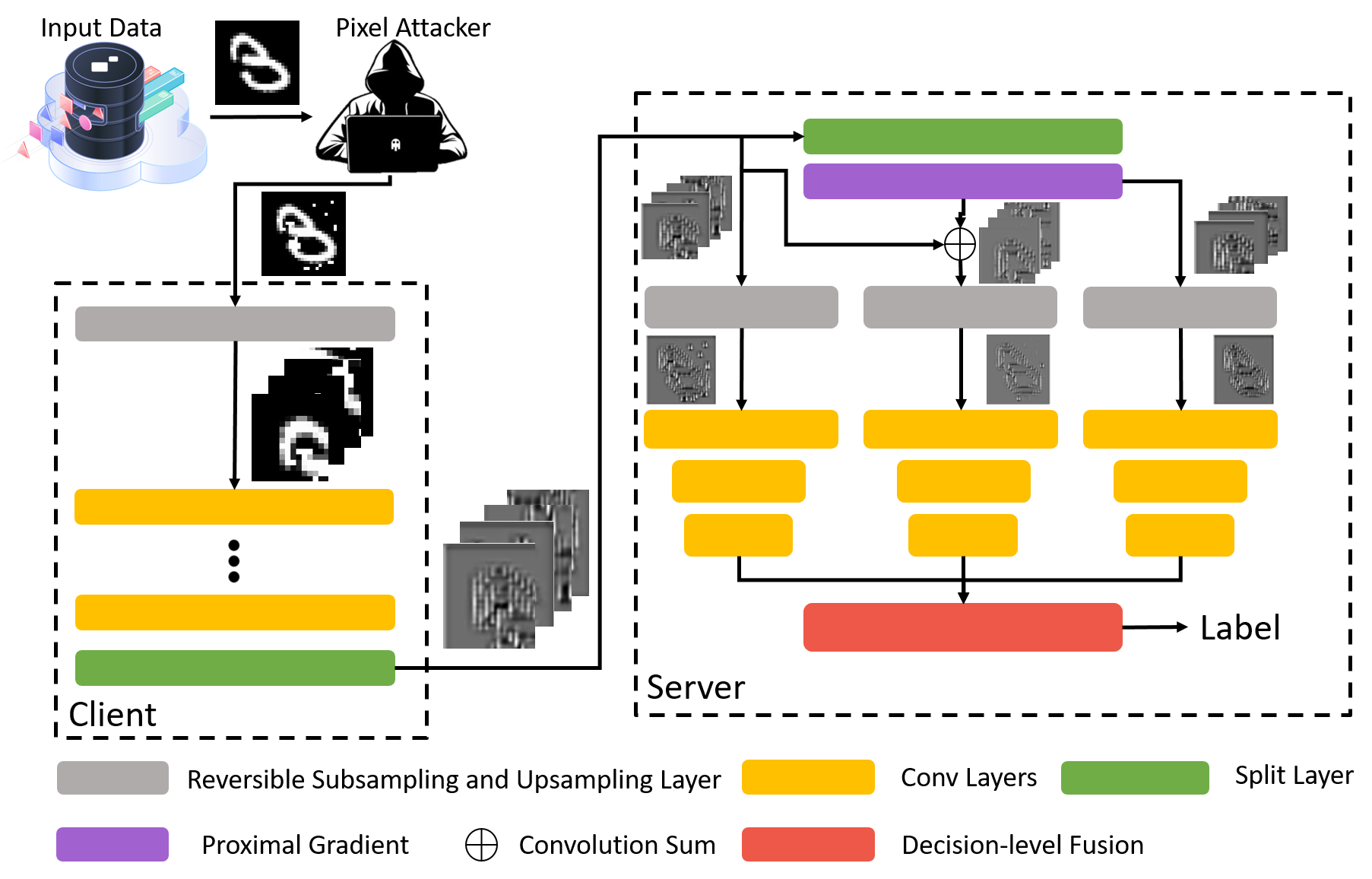}
\caption{Workflow of Proximal Gradient Split Learning network}
\label{fig:2}
\end{figure*}

The workflow of PGSL method is illustrated in Figure \ref{fig:2}. An example from MNIST dataset is used to illustrate the process due to visual clarity, however, the same process is applied to the IoMT when used with publicly available medical dataset in this study. An intentional pixel attack is initiated towards the IoMT data, which then undergoes a sub-sampling layer that divides the medical image into patches. These patches in the form of a tensor are sent to the CNN for partially training the network at the client side. The network is segmented at the split/cut layer along with the extraction of outputs. These outputs are then sent to server side and processed through the proximal gradient method to remove the pixel attacks. We branch out three streams from this point, the first one retains the up-sampled pixel-attacked data, the second performs a convolution sum between the pixel-attacked and proximal gradient data, and the third uses only proximal gradient data. The streams are trained at the server side using forward propagation. The gradients from the last layer of the second stream (combining both the pixel-attacked and proximal gradient data) at the server side are backpropagated to the last split layer. Only these gradients are sent back to the client side to fine-tune the training process. Once the network is trained, we employ a weighted-averaging decision-level fusion method to derive the desired label. The details for each of the PGSL building blocks are provided in the subsequent subsections. In the above scenario, the client and server side can be considered analogous to patient and hospital, clinic and laboratories, university labs and medical research institutes, respectively.

\subsection{Pixel-attack on IoMT data}
In this study, we mainly consider the image data for designing the privacy-preserving machine learning (PPML) method. For this study, we use jacobian-based saliency map attack (JSMA) pixel method \cite{JSMA} to initiate the attack. The rationale for choosing JSMA method is two-fold. The first is the convenience to implement such an attack as it mostly performs correlation and the second is the effectiveness of JSMA approach. It has been revealed later in the experimental results (see Table II) that the data attacked using JSMA method is more difficult to recover from activation maps in comparison to other attack methods, respectively. Let us consider that an image label pair is represented as $(x,y)$. The saliency map observes the influence of each pixel in image $x$ for predicting the class $y$. The assumption is that the pixel correlates to the corresponding class positively $Corr_p = \frac{\partial f(x)_{y}}{\partial x_i}>0$ and to the contradicting class negatively $Corr_n = \sum_{y'\neq y} \frac{\partial f(x)_{y'}}{\partial x_i} < 0$, where $f(x)$ refers to the softmax probabilities and $y'$ corresponds to contradicting classes. Based on the aforementioned assumptions, the map can be formulated as shown in equation 1.
\begin{equation}
  Map = \begin{cases} -\frac{\partial f(x)_{y}}{\partial x_i} \cdot \sum_{y'\neq y} \frac{\partial f(x)_{y'}}{\partial x_i}, & \text{if } True\\ 0,          &\text{otherwise} \end{cases}
\end{equation}
The condition $(\emph{True})$ in equation 1 refers to the satisfaction of $Corr_p$ and $Corr_n$ conditions. The attacker can target the saliency map such that the pixels are modified to increase the correlation of contradictory label. One simple way is to inverse the correlation conditions, i.e. $Corr'_p = \frac{\partial f(x)_{y}}{\partial x_i}<0$ and $Corr'_n = \sum_{y'\neq y} \frac{\partial f(x)_{y'}}{\partial x_i} > 0$. The process of generating adversarial sample in the aforementioned way is best suited to this study as we do not intent to initiate a targeted attack but rather a non-targeted one. Let us denote the adversarial saliency map as $Map'$, therefore the resultant attacked image can be given by $\hat{x} = x + Map'$

\subsection{Reversible Up-sampling and Down-sampling Layer}
Post the initialized attack, we employ a reversible down-sampling and up-sampling layer \cite{Khowaja2021}, accordingly. The down-sampling is applied at the client side, while the up-sampling is performed at the server side. The reason for using this layer is three-fold. Firstly, the attacker needs a prior information for design of sub-pixel convolution to up-sample the image. Secondly, the increasing the receptive field while retaining the depth, and thirdly, the reduction of artifacts that could affect the visual quality \cite{Khowaja2021}. Suggesting that the size of adversarial image $\hat{x}$ is represented by $i \times j$, it will be down-sampled into four images along with the concatenation of corresponding saliency map $Map$ to form an input tensor having size $\frac{i}{2} \times \frac{j}{2} \times (4ch+1)$ where $ch$ corresponds to the number of channels, i.e. 1 for grayscale and 3 for color. Let us denote the tensor as $\breve{x}$. 

\subsection{Proximal Gradient for IoMT data}
Let us represent the feature maps with $\Grave{\Grave{x}}$ that are driven from $\breve{x}$, i.e. $\breve{x} \overset{h}{\rightarrow} \Grave{\Grave{x}}$, where $h$ represents the convolution function used to extract the activation. These activations are passed as an input to the proximal gradient method to recover the attacked IoMT data, accordingly. Although we adopt the proximal gradient method (which is a broad family of functions that optimizes convex gradient methods), it is not the same. The method has been modified in terms of the modalities, i.e. we use corrupted data with maps instead of actual and modified images, and in terms of equality constraint to comply with the proposed network architecture.\\
Recalling the initiated attack, the IoMT data can be recovered by $x = \hat{x} - Map'$, however, the information regarding the $Map'$ is not available in the activations. Therefore, we approximate the recovery using the inverted saliency map concatenated in the tensor activation, i.e. $x \approx \hat{x} - \bar{Map}$ and represent it as an optimization function shown in equation 2.
\begin{equation}
    \displaystyle \min_{\hat{x},\bar{Map}} \Vert \hat{x} \Vert_* - \lambda \Vert \bar{Map} \Vert_*
\end{equation}
Equation 2 is considered to be a convex optimization problem \cite{Canyi2016}. The notation $\Vert \Vert_*$ refer to the nuclear norm of the matrix and $\lambda$ represents the weighting parameter, accordingly. The said equation is also referred to as robust principle component analysis \cite{PG2009, Helmut2019} which is commonly used for image recovery. Equation 2 is also considered to be a special case of a general optimization problem that can be represented in the form shown in equation 3.
\begin{equation}
    \displaystyle \min_{\chi \in \mathcal{H}} g(\chi), \text{s.t. }  \mathcal{A}(\chi) - b = 0
\end{equation}
The notation $g, \mathcal{H}, \mathcal{A}$ and $b$ refer to the convex function, real Hilbert space, linear map, and an observation, respectively. An efficient way to solve equation 3 is to relax the equality constraint and represent it into the following form.
\begin{equation}
    \displaystyle \min_{\chi \in \mathcal{H}} \mathcal{F}(\chi) \doteq \mu g(\chi) + f(\chi)
\end{equation}
In the context of this study,
\begin{equation}
    f(\chi) \doteq \frac{1}{2} \Vert \hat{x} - \bar{Map} \Vert^2
\end{equation}
that is responsible for penalizing in case of equality constraint violation, $g$ is the convex function subject to $\hat{x} \rightarrow Map$, and the $\mu$ corresponds to the relaxation parameter subject to $\mu > 0$. The assumption is that the solution of equation 3 approaches to that of equation 2 as the $\mu$ approaches 0. The function in equation 5 is assumed to be smooth and convex, thus, it can be solved by using Lipschitz continuous gradient function \cite{PG2009, Helmut2019} shown in equation 6.
\begin{equation}
    \Vert \nabla f(\chi_1) - \nabla f(\bar{Map}_1) \Vert \leq \mathcal{L} \Vert \chi_1 - \bar{Map}_1 \Vert
\end{equation}
The Fr$\acute{e}$chet derivate is represented by $\nabla f$ that is represented as an element in the Hilbert space. The use of Lipschitz function has proven to make the solution more efficient in terms of computational complexity. The optimization function shown in equation 4, correspond to the family of proximal gradient function that is used to recover the attacked image in this study. 

\subsection{Fusion of Activation Maps}
As illustrated in Figure \ref{fig:2}, we perform the fusion of output activation maps from the split layer and the proximal gradient method. There are various ways to fuse the activation maps, but the most common ones are sum, convolution, and convolution-sum fusion strategies. It has been proven in existing studies that the convolution-sum fusion yields better results in comparison to the former ones. In this regard, we adopt the convolution-sum fusion strategy proposed in \cite{KhowajaNCAA2020} to fuse the activation maps, accordingly. The fusion comprises the orderly steps such as concatenation, convolution, dimension reduction, and summation. For simplicity, we represent the feature map with $\hat{f}$, the two feature maps that needs to be fused are denoted as $\hat{f}_{u}$ and $\hat{f}_{v}$, respectively. The steps for performing the fusion are given below. 
\begin{itemize}
    \item The first step concatenates the activation maps at some spatial locations across the channels.
    \item A bank of filters and biases are used to perform the convolution in the second step.
    \item The third step performs the dimensionality reduction within the convolution process by generating a weighted combination of the activation maps.
    \item The last step performs a linear summation of the corresponding maps that needs to be fused.
\end{itemize}
We represent the mathematical formulation for the fusion of feature maps as shown in equation 7 and 8.
\begin{equation}
    \hat{f}_{out} = sum(conv(\hat{f}_{u}, \hat{f}_{v}), \hat{f}_{u})
\end{equation}
\begin{equation}
    \hat{f}_{out} = (concat*filt + bias) + \hat{f}_{u}
\end{equation}
The first and second steps corresponds to $conv$ operation that employs a bank of filters $filt$ and biases $bias$, as represented in equation 7 and 8. The $conv$ operation further reduces the dimension so that the summation operation can be performed, accordingly. The last step performs a linear summation $sum$ between the output of convolution operation and $\hat{f}_{u}$.

From this point forward, three streams are trained using the attacked image, fused image, and the recovered image. 

\subsection{Decision-Level fusion}
Existing studies have proven that defense measures for adversarial and model inversion attacks heavily affect the recognition performance of the system. In this regard, PGSL employs a decision-level fusion strategy that combines the classification results from the three streams. Generally, three kinds of fusion strategies, i.e. Weighted-Averaging, Adaptive-Weighted-averaging, and meta-learning, are employed to improve the recognition performance \cite{KhowajaNCAA2020}. On one hand, meta-learning is considered to be more effective while being computationally complex and on the opposite spectrum, weighted-averaging is simple and has  the least computational constraints. Adaptive-weighted-averaging provides an efficient trade-off between the effectiveness and computational complexity \cite{KhowajaNCAA2020}, thus in this work, we use adaptive-weighted-averaging for decision-level fusion. Let us denote the classification scores from attacked image stream, fused activation maps stream, and recovered image stream as $\mathcal{S}_a, \mathcal{S}_f,$ and $\mathcal{S}_r$, respectively. The adaptive-weighted-average for combining the scores from aforementioned three streams can be defined as shown in equation 9.
\begin{equation}
    \mathcal{S}_{awa} = \gamma * \mathcal{S}_r + \rho * \mathcal{S}_f + (1- \gamma - \rho)* \mathcal{S}_a 
\end{equation}
where $\gamma$ and $\rho$ represent the weights for scores from recovered and fused activation maps, respectively. Let us denote the corresponding weights for the three streams as $\mathbb{W}_{\gamma}$, $\mathbb{W}_{\rho}$, and $\mathbb{W}_{\beta}$. We first initialize the fixed weights as describe in experiments section and compute the values of $\gamma$ and $\rho$ as shown in equation 10 and 11.
\begin{equation}
    \gamma = \frac{\mathbb{W}_{\gamma} * \mathcal{S}_r^{max}}{\mathbb{W}_{\gamma} * \mathcal{S}_r^{max} + \mathbb{W}_{\rho} * \mathcal{S}_f^{max} + \mathbb{W}_{\beta} * \mathcal{S}_a^{max}}
\end{equation}
\begin{equation}
     \rho = \frac{\mathbb{W}_{\rho} * \mathcal{S}_f^{max}}{\mathbb{W}_{\gamma} * \mathcal{S}_r^{max} + \mathbb{W}_{\rho} * \mathcal{S}_f^{max} + \mathbb{W}_{\beta} * \mathcal{S}_a^{max}}
\end{equation}
where $\mathcal{S}^{max}$ represent the maximum average score of a particular class label and can be defined for the corresponding streams as $\mathcal{S}_r^{max} = \max_{\mathbb{L}}[\mathcal{S}_r(\mathbb{L})]$,  $\mathcal{S}_f^{max} = \max_{\mathbb{L}}[\mathcal{S}_f(\mathbb{L})]$, and $\mathcal{S}_a^{max} = \max_{\mathbb{L}}[\mathcal{S}_a(\mathbb{L})]$. The notation $\mathbb{L}$ represents the class label. 
\subsection{Network Configuration}
As the scope of this work is to demonstrate the effectiveness of defense against model inversion attack and data recovery to improve the recognition performance, we adopt a simple convolutional neural network (CNN) with 7 conv and 2 fully connected (FC) layers that can be used for the employed datasets. Each of the convolutional layer comprises conv, ReLU, and batch normalization (BN) layer with 3x3 kernel size and 64 channels. The drop-out layer is used after every three conv layers. We employ the split layer after 2nd conv layer, accordingly. The reason for choosing 2nd conv layer for the split is driven by the research findings from \cite{Wu2021, He2019} that the reconstruction of intermediary outputs work better when performed at initial layers as the structural similarity of the feature maps is high with the input data. Furthermore, we also tried splitting the network at other layers but achieved the best results with the proposed settings, therefore, we assume that our empirical finding is compliant with the existing works for performing split at initial layers, respectively. The details regarding the hyperparameters and distribution of datasets is given in experiment section. 

\section{Experimental Setup and Analysis}\label{sec:exp}
This section provides the experimental setup, results, and analysis to show the effectiveness on two fronts. The first is the defense mechanism for reconstruction of images from activation maps and its recovery, and the second is the recognition performance. We present extensive experimental analysis to show the effectiveness of the proposed approach. The corresponding results for each of the component is shown in subsequent subsections.

\subsection{Experimental Setup}
As the study is centered around IoMT data, we employ two datasets to prove the efficacy of PGSL. The first is a publicly available Mammogram dataset MIAS \cite{Suckling1994} and the second is the MNIST dataset \cite{Lecun1998}. The rationale for choosing MNIST dataset is the fairness of comparison with existing approaches and clarity of visual results. There are a total of 330 images in MIAS dataset. We clip the images to 1024x1024 and divide them into training and testing sets. The training set comprises 42, 57, and 181 while the testing set contains 12, 12, and 26 malignant, benign, and normal images, respectively. For MNIST, the training and testing set comprises 60,000 and 10,000 images. The network for both the datasets employ same set of parameters. We use the learning rate of 0.001 with a decay rate of 0.0003, the drop out ratio is 0.25, and the optimizer is set to ADAM. All the experiments are performed on Python, Intel Core i9 PC clocked at 3.5 GHz with 64GB of RAM and NVIDIA GeForce RTX3090.

\subsection{Experiments with varying $\mu$}
\begin{figure}[!htbp] %!t
\centering
\includegraphics[width=\linewidth]{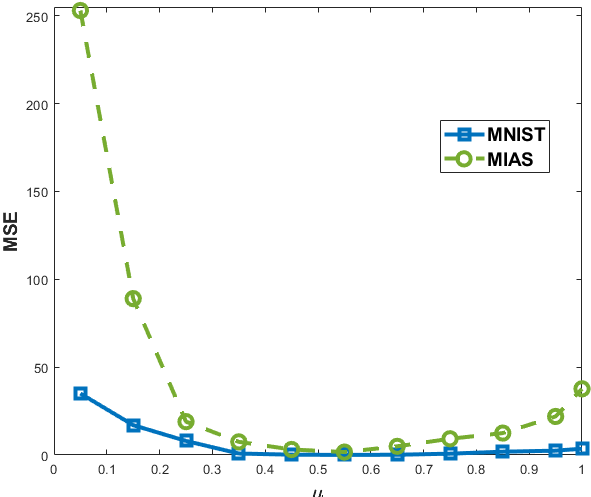}
\caption{Sensitivity analysis for parameter $\mu$ on MNIST and MIAS datasets}
\label{fig:3}
\end{figure}

The proximal gradient method in this study relies on a hyperparameter, i.e. $\mu$, for the optimal recovery of the attacked image. The mean squared error (MSE) is computed between the recovered and the original image (before attack initiation) to select the optimal $\mu$ value. Since the value of $\mu = 0$ will yield the same result as of equation 2, we start the selection of value from 0.05 to 1.0 with the step size of 0.1. We conducted these experiments on the attacked image before giving it as an input to downsampling layer or CNN. The reason for not conducting on the activation maps is that the input is an attacked image therefore, comparing the corresponding activation maps would not be meaningful. We measure the MSE against the $\mu$ values on both the datasets. The results for this experiment are shown in Figure \ref{fig:3}. For the sake of generality, we choose a single value of $\mu$ for both the datasets. The analysis indicate that $\mu=0.55$ yields the lowest MSE for both the datasets, therefore, we will use this value for our next set of experiments, accordingly.  

\subsection{Comparative Analysis for Reconstruction of Attacked Data}
\begin{figure}[!htbp] %!t
\centering
\includegraphics[width=\linewidth]{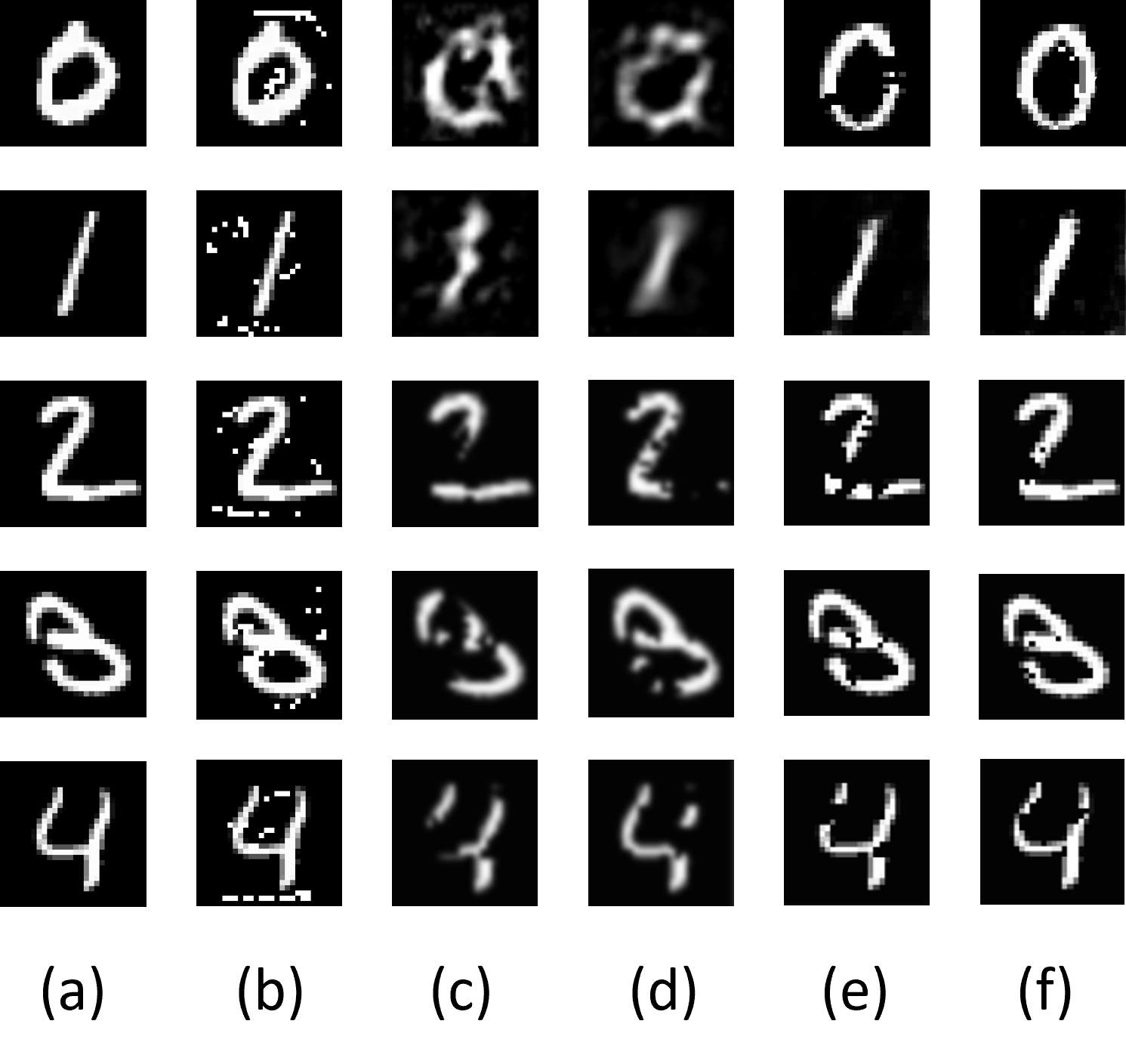}
\caption{Qualitative results on MNIST dataset (a) Original image, (b) Attacked image, (c) Recovered image using DCGAN, (d) Recovered image using DLG, (e) Recovered image using DCGAN+Proximal Gradient, and (f) Recovered image using DLG+Proximal Gradient}
\label{fig:4}
\end{figure}

To show the effectiveness of the proposed approach in terms of the reconstruction of attacked data, we use state-of-the-art methods to recover images from activation maps, i.e. deep leakage for gradient (DLG) \cite{Zhu2020}, improved deep leakage from gradient (iDLG) \cite{Zhao2020}, and DCGAN \cite{Hitaj2017} method to reconstruct the images from their gradients. The DLG and iDLG method already provides a pre-trained network for MNIST dataset, but we trained the DLG network for MIAS dataset in order to obtain the recovered images from random initialization. Similarly, DCGAN was also trained from random initialization in order to recover the images from gradients. We first consider the gradients from the split layer and then the gradients from proximal gradient method to recover the images to prove the efficacy of the proposed approach. We evaluate the method using MSE, as the comparison with existing approaches would be fair enough. The visual results for DLG and DCGAN on MNIST without and with proximal gradient is shown in Figure \ref{fig:4}. We also present the quantitative results for DLG, iDLG, and DCGAN in terms of MSE on both the datasets in Table 1. It can be deduced from the results that the JSMA attack method used in this study is more difficult to recover from gradient/activation maps, thus, we assume that the PGSL framework has a better defense concerning model inversion attacks. For recovery method, we corrupted the images with laplacian noise \cite{Titcombe2021} and used DLG, iDLG, and DGCAN to recover the images. The results show that the proposed PGSL method is able to improve the MSE concerning recovered images. 

\begin{table}[]
\centering
\caption{Comparative Analysis of existing works using MSE on MIAS and MNIST with proposed Attack and Recovery Methods}
\label{tab:1}
\begin{tabular}{|c|c|c|}
\hline
Attack Method     & MIAS  & MNIST \\ \hline
DLG \cite{Zhu2020}         & 3.14  & 0.12  \\ \hline
iDLG \cite{Zhao2020}         & 2.63  & 0.048  \\ \hline
DCGAN \cite{Hitaj2017}         & 2.96  & 0.056 \\ \hline
MIA \cite{Wu2021}         & 2.925 & 0.08  \\ \hline
JSMA + DLG    & 4.362 & 0.24  \\ \hline
JSMA + iDLG    & 3.924 & 0.22  \\ \hline
JSMA + DCGAN    & 5.947 & 0.29  \\ \hline
Recovery Method   & MIAS  & MNIST \\ \hline
\cite{Titcombe2021} + DLG   & 2.543 & 0.097 \\ \hline
\cite{Titcombe2021} + iDLG   & 1.349 & 0.04 \\ \hline
\cite{Titcombe2021} + DGCAN & 7.86  & 2.34  \\ \hline
Ours (DLG)        & 1.854 & 0.046 \\ \hline
Ours (iDLG)        & 1.126 & 0.026 \\ \hline
Ours (DGCAN)      & 2.372 & 0.078 \\ \hline
\end{tabular}
\end{table}

\subsection{Experimental Results on Recognition Performance}
\begin{figure}[!htbp] %!t
\centering
\includegraphics[width=0.85\linewidth]{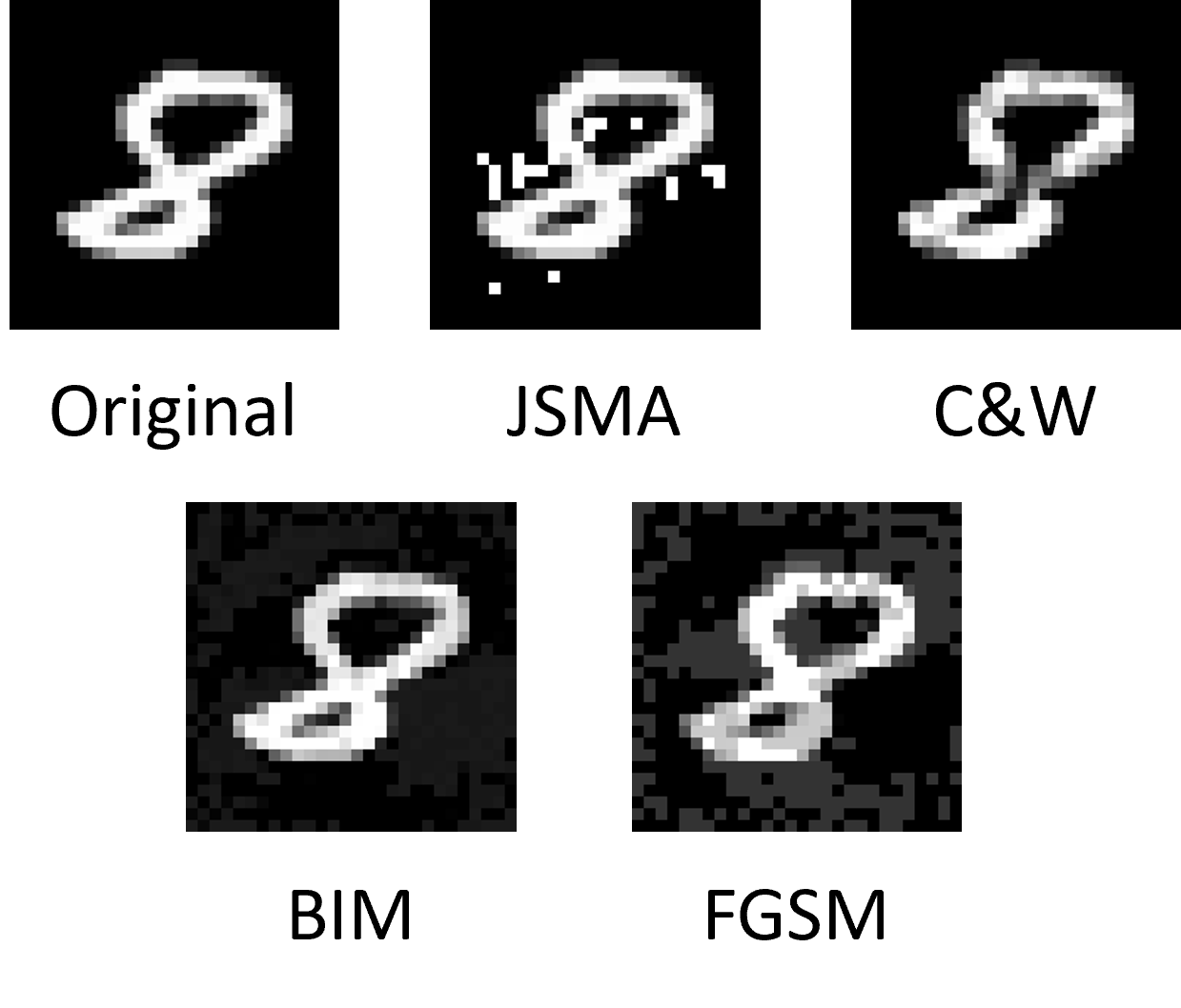}
\caption{Visual results of different attacks on images}
\label{fig:5}
\end{figure}

We present an experimental analysis to show the effectiveness of PGSL in terms of recognition performance. For this experiment, we test the recognition accuracy on the images if attacked using the adopted jacobian method \cite{JSMA}, carlini and wagner (C\&W) method \cite{carlini2017}, basic iterative method (BIM) \cite{BIM2017}, and fast gradient sign method (FGSM) \cite{Goodfellow2014}, respectively. The attacked images using the respective methods are shown in Figure \ref{fig:5}. These attacks have proven to be effective for making the end predictions completely or partially wrong, as illustrated in the visual results. We report the results when trained directly with the attacked images, the images constructed using DLG, iDLG, and DCGAN from maps acquired using split layer, and the proposed PGSL method in Table 2. The results indicate that the JSMA is a highly effective attack method when it comes to MIAS while C\&W attack yields the lowest accuracy on MNIST. The BIM and FGSM method are weaker attacks relative to the JSMA and C\&W, however, they do affect the overall accuracy of the recognition system. iDLG relatively performs better than DLG and DGCAN. The stream computed on images recovered from proximal gradient yields the best accuracy, better than iDLG, while the other two streams yield lower results. Considering that the attack is incorporated in the other two streams directly or indirectly, the degradation of performance makes sense. In this regard, we used the decision-level fusion strategy using adaptive-weighted-averaging method. We initialized the weights for all the streams with 0.5, 0.3, and 0.2, respectively, based on the results from individual streams and applied the fusion to derive the final label. The purpose of using the fusion of decisions from multiple streams is not only to improve the recognition performance but also to make the recognition network attack resilient, which is supported by the highest recognition accuracy achieved on both the datasets. It should be noted that the accuracies may vary from the existing works, as we trained and tested the images using the proposed CNN network. 
\begin{table}[]
\centering
\caption{Comparative Analysis on Accuracies for MIAS and MNIST Datasets}
\label{tab:2}
\begin{tabular}{|c|c|c|}
\hline
Method          & MIAS   & MNIST  \\ \hline
JSMA \cite{JSMA}            & 46.3\% & 65.4\% \\ \hline
C\&W \cite{carlini2017}           & 51.2\% & 58.6\% \\ \hline
BIM \cite{BIM2017}            & 56.8\% & 81.2\% \\ \hline
FGSM \cite{Goodfellow2014}           & 61.9\% & 83.8\% \\ \hline
DLG \cite{Zhu2020}            & 65.7\% & 88.3\% \\ \hline
iDLG \cite{Zhao2020}            & 69.2\% & 95.4\% \\ \hline
DCGAN \cite{Hitaj2017}          & 62.2\% & 84.8\% \\ \hline
$\mathcal{S}_r$ & 78.5\% & 99.2\% \\ \hline
$\mathcal{S}_f$ & 67.3\% & 86.6\% \\ \hline
$\mathcal{S}_a$ & 46.3\% & 65.4\% \\ \hline
PGSL            & 83.2\% & 99.8\% \\ \hline
\end{tabular}
\end{table}
\section{Discussion and limitations} \label{sec:disc}
Referring to Figure \ref{fig:1}, we illustrated different types of attacks that can applied to IoMT data that can result in severe consequences such as data theft, wrong diagnosis, financial losses, and more. The first attack refers to the raw IoMT data which is attacked while being sent to a hospital in a use-case scenario. The results in Table 2 clearly indicates that any of the attack is capable of reducing the accuracy to almost a level where it's a little better than taking a random decision. Model inversion attacks are also severe as they are able to reconstruct the raw data from gradient/activation maps, which violates the privacy of a patient/user. The PGSL framework shows that it can provide some defense concerning the information associated with the raw data. iDLG is considered to be a state-of-the-art method for reconstruction of images from gradient maps, but the results show that the proposed work helps in reducing recognition performance from the reconstructed images, thus, by extension, reduces the degree of recovery from gradients maps. The strength of the proposed work lies within its adoption in several emerging domains such as Spatial Computing, Virtual Medicine, Digital Twins, and Metaverse. All the aforementioned domains are concerned with simulating humans in the digital world. The proposed work could be greatly helpful for preserving the users' data if any of the aforementioned technologies is realized for IoMT ecosystem.  \\
Although the PGSL method serve its purpose, it assumes the $Map$ to be available from the data aggregator stage, however, with current progress in GANs, it has the capability to evolve for such defense mechanism. Furthermore, considering that the IoMT data is highly sensitive, and a slight perturbation can cause the wrong diagnosis, the achieved accuracy still has the room for improvement when it comes to IoMT data. Nevertheless, PGSL reports approximately, 36.9$\%$, 17.5$\%$, and 14$\%$  gains in comparison to the JSMA, DLG, and iDLG methods, accordingly. Moreover, this study considers mammogram images as IoMT data, but there are other homogeneous and heterogeneous medical modalities that can be explored for such adversarial affects such as X-Ray images, Fundus images, CT-scans, medical reports, electronic health records, and so forth. 
\section{Conclusion and Future Works}\label{sec:con}
In this study, we have proposed PGSL framework for defense against the impact of data theft and model inversion attacks within an IoMT ecosystem. The underlying idea of PGSL shows that it not only helps in manipulating attacker for having the wrong or partial information from the IoMT data, but also helps in defending the information against model inversion attacks. We also show through our analysis that the PGSL method can be used with other techniques to improve both the recovery and recognition performance, accordingly. The method has been tested on MIAS and MNIST dataset that proves the effectiveness of the proposed approach. The implication of PGSL can easily be realized in any IoMT ecosystem, ranging from e-health to spatial computing domains. Furthermore, the PGSL can be applied to multiple datasets, in general, provided that the hyperparameters are fine-tuned for that particular dataset (specifically the data modality), accordingly. \\
There are several directions in which the current work can be extended. One of the possible directions is its use in Private AI framework that can help in securing both data and model security. Another future work is to observe the effect of data privacy preservation when it comes to the adoption of virtual worlds such as Metaverse, Spatial computing, and Digital Twins. Lastly, the proposed work can be extended to observe its effect on multiple client nodes or server nodes with split learning concerning the domain adaptation strategy. 

\section*{Acknowledgement}
This research was supported by Basic Science Research Program through the National Research Foundation of Korea (NRF) funded by the Ministry of Education (2017R1A6A1A03015562).

\bibliography{ref.bib}
%%%%%%%%%%%%%%%%%%%%%%%%
\bibliographystyle{IEEEtran}
%\bibliography{Biblio}
%%%%%%%%%%%%%%%%%%%%%%%%
\vspace{-0.5cm}
%%%%%%%%%%%%%%%%%%%%%%%%

% % %%%%%%%%%%%%%%%%%%%%%%%%

\begin{IEEEbiography}[{\includegraphics[width=1in,height=1.25in,clip,keepaspectratio]{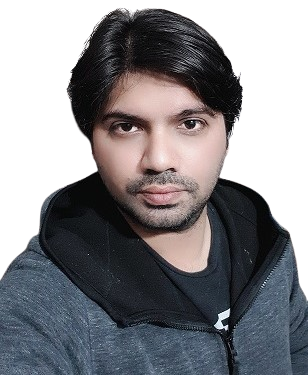}}]{Sunder Ali Khowaja}
received the Ph.D. degree in Industrial and Information Systems Engineering from Hankuk University of Foreign Studies, South Korea. He has also completed his postdoctoral research fellowship at Department of Mechatronics Engineering, Tech University of Korea. He is currently serving as an Assistant Professor at Department of Telecommunication Engineering, Faculty of Engineering and Technology, University of Sindh, Pakistan. He had the experience of working with multinational companies as Network and RF engineer from 2008-2011. He has teaching, research, and administrative experience of more than 12 years. He has published over 40 research articles in International Journals, and Conference Proceedings. He is also a regular reviewer of notable journals from IEEE Transactions, IET, Elsevier, and Springer. He has also served as a TPC member for workshops in A* conferences such as CCNC 2021, Globecom 2021, and MobiCom 2021. In addition, he is serving as a guest editor for special issues in Computers and Electrical Engineering, Human-Centric Computing and Information Sciences, Sustainable Energy Assessments and Technologies, and Journal of King Saud University: Computing and Information Sciences. His research interest includes Deep Learning, Data Analytics, Computer Vision, Model Security, and Private AI. 
\end{IEEEbiography}
% % %%%%%%%%%%%%%%%%%%%%%%%%

\vskip -2\baselineskip plus -1fil
\vspace{-0.25cm}
% % %%%%%%%%%%%%%%%%%%%%%%%%

\begin{IEEEbiography}[{\includegraphics[width=1in,height=1.25in,clip,keepaspectratio]{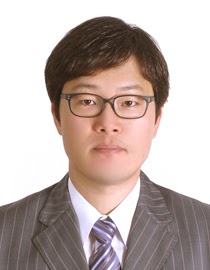}}]{Ik Hyun Lee} received the B.S. degree in control and instrument engineering from Korea University, South Korea, in 2004, and the M.S. and Ph.D. degrees from the School of Information and Mechatronics, Gwangju Institute of Science and Technology, South Korea, in 2008 and 2013, respectively. He was a Postdoctoral Researcher at the Media Laboratory, Massachusetts Institute of Technology, and a Senior Researcher at the Korea Aerospace Institute of Research. He is currently an Assistant Professor with the Department of Mechatronics Engineering and Korea Polytechnic University, South Korea. His research interests include image registration, image fusion, depth estimation, and medical image processing. 
\end{IEEEbiography}
% % %%%%%%%%%%%%%%%%%%%%%%%%

% % %%%%%%%%%%%%%%%%%%%%%%%%

\vskip -2\baselineskip plus -1fil
\vspace{-0.25cm}

\begin{IEEEbiography}[{\includegraphics[width=1in,height=1.25in,clip,keepaspectratio]{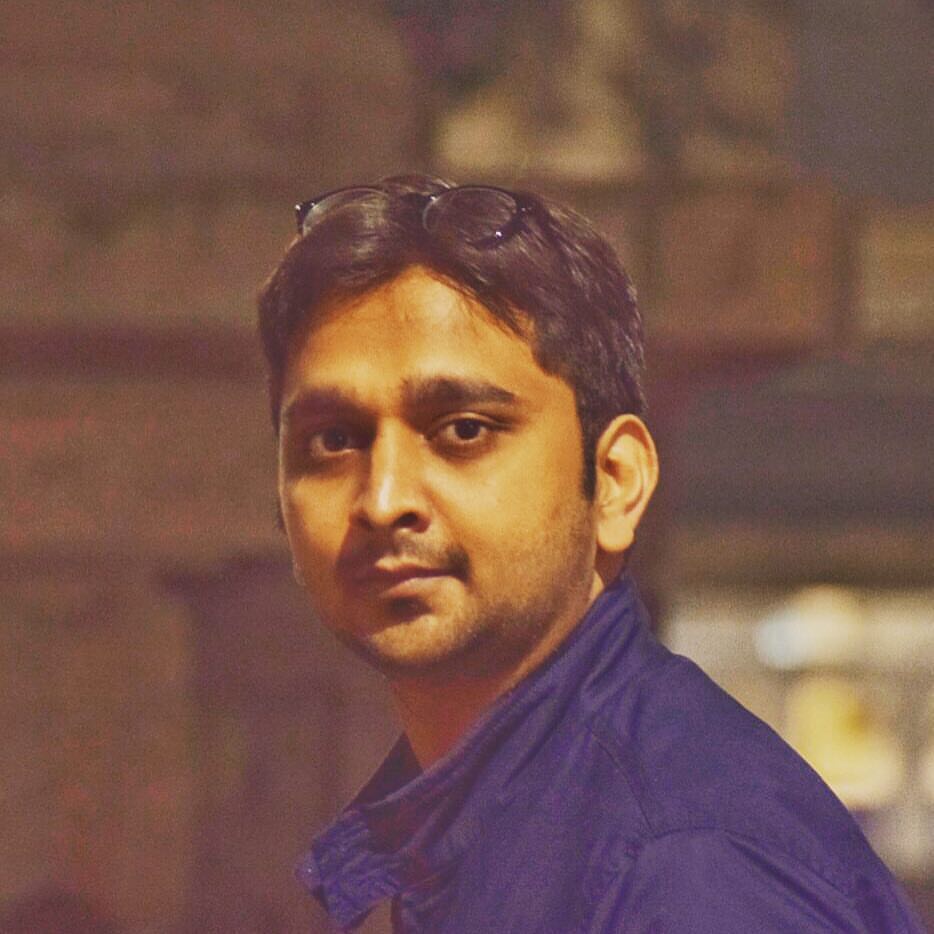}}]{Kapal Dev} is currently serving as Assistant Lecturer at Munster Technological University (MTU), Ireland and formerly he was senior researcher at same University. Previously, he was a Postdoctoral Research Fellow with the CONNECT Centre, Trinity College Dublin (TCD). He is very active in leading (as Principle Investigator) Erasmus + International Credit Mobility (ICM), Capacity Building for Higher Education, and H2020 Co-Fund projects and won 0.45 million Euros funding in total as well.  He is serving as Associate Investigator (AI) by CONNECT Centre, Trinity College Dublin (TCD) funded by Science Foundation. He worked for OCEANS Network as Head of Projects funded by European Commission. He is founding chair of IEEE ComSoc special interested group titled as Industrial Communication Networks. He is serving as Associate Editor in NATURE, Scientific Reports, Springer WINE, HCIS, IET Quantum Communication, IET Networks, Area Editor in Elsevier PHYCOM. He performed duties as Guest Editor (GE) in IEEE Network, IEEE TII, IEEE TNSE, IEEE TGCN, IEEE STDCOMM. He has published over 40 research papers majorly in top IEEE Transactions, Magazines and Conferences.He is an expert external evaluator of European Research Council (ERC) starting grant, several MSCA Co-Fund schemes, Elsevier, IET, Springer Book proposals and top scientific journals and conferences.  His research interests include Wireless Communication Networks, Blockchain and Artificial Intelligence.
\end{IEEEbiography}
% % %%%%%%%%%%%%%%%%%%%%%%%%

\vskip -2\baselineskip plus -1fil
\vspace{-0.25cm}

\begin{IEEEbiography}[{\includegraphics[width=1in,height=1.25in,clip,keepaspectratio]{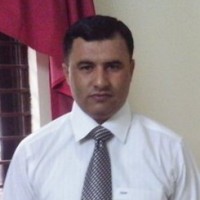}}]{Muhammad Aslam Jarwar}
is a Senior Lecturer at the Department of Computing, Sheffield Hallam University, UK and co-investigator for the Secure Ontologies for IoT Systems (SOfIoTS) project. He is an Associate Fellow of the Higher Education Academy (AFHEA), UK, Senior Member IEEE and a private member of the 4-Dimensionalism Special Interest Group (4DSIG) Network. Previously, he served as a Research Fellow at University College London (UCL), UK and Postdoctoral Research Associate at the University of Manchester, UK. He received the Ph.D. in Information and Communications Engineering from Hankuk University of Foreign Studies, South Korea. Currently, he is serving as a lead guest editor for the special issue in Elsevier journal on Sustainable Energy Technologies and Assessments, editorial board member of Progress in Human Computer Interaction journal and TPC for the international conference on sustainable technologies for industry 4.0.  He is the author of several peer-reviewed articles and ITU-T technical reports. In addition, he is a regular reviewer of IEEE Transactions, Elsevier and Springer Q1 journals. His research interest includes IoT/IIoT, IoT devices security, Digital Twin, Deep Learning, Ontologies and AI.
\end{IEEEbiography}

\vskip -2\baselineskip plus -1fil
\vspace{-0.25cm}
\begin{IEEEbiography}[{\includegraphics[width=1in,height=1.25in,clip,keepaspectratio]{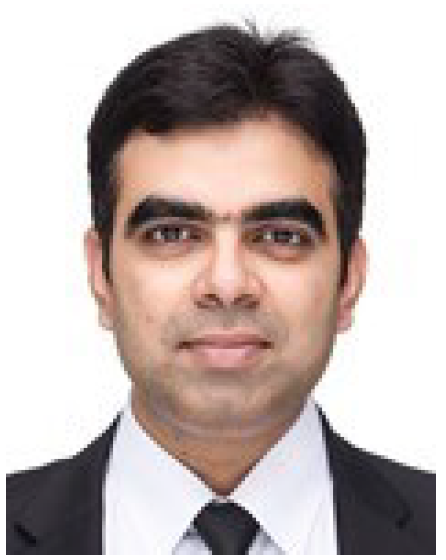}}]{Nawab Muhammad Faseeh Qureshi} is as assistant professor in the Department of Computer Education in Sungkyunkwan University, Republic of Korea. He was awarded the 1st Superior Research Award from the College of Information and Communication Engineering based on his research contributions and performance. He has served as Guest Editor in more than 19 Journals. He has graduated several masters student and a Ph.D. student so far. He has published several papers in TOP tier journals such as IEEE Internet of Things, IEEE Transactions on Industrial Informatics, Sustainable Cities and Society, IEEE Transactions on Intelligent Transportation Systems, IEEE Transactions on Fuzzy Systems and Flagship conferences such as IEEE global communications conference (GLOBECOM), IEEE International Conference on Distributed Computing Systems (ICDCS) and others. He is a Senior Member IEEE and actively participates in online Webinars and has given several keynotes in the conferences and Seminars. His area of research is convergence of big data with machine learning, deep learning, context-aware processing in IoT devices and their communication protocols for 5G and onwards.
\end{IEEEbiography}
\vskip -2\baselineskip plus -1fil
\vspace{-0.25cm}

\end{document}